\documentclass[preprint]{revtex4}
\usepackage{graphicx}
\usepackage{latexsym}
\usepackage{amsmath}
\newcommand{\beq}{\begin{equation}}
\newcommand{\eeq}{\end{equation}}
\newcommand{\beqa}{\begin{eqnarray}}
\newcommand{\eeqa}{\end{eqnarray}}
\newcommand{\bseq}{\begin{subequations}}
\newcommand{\eseq}{\end{subequations}}
\newcommand{\bold}{\boldsymbol}
\newcommand{\wtilde}{\widetilde}

\newcommand{\trm}{\textrm}

\newcommand{\id}{{1\!\!1}}

\begin{document}

\title{Screening Masses of Scalar and Pseudo-scalar Excitations in Quark-gluon Plasma}
\author{P. Czerski\email{piotr.czerski@ifj.edu.pl}}
\affiliation{The H. Niewodnicza\'nski Institute of Nuclear Physics, \\ Polish Academy of Sciences,\\
     ul. Radzikowskiego 152, PL-31-342 Krak\'ow, Poland}

\author{W.M. Alberico}
\affiliation{Dipartimento di Fisica dell'Universit\`a di Torino and \\
  Istituto Nazionale di Fisica Nucleare, Sezione di Torino, \\
  via P.Giuria 1, I-10125 Torino, Italy}

\begin{abstract}
The quark-gluon plasma (QGP) excitations, corresponding to the 
scalar and pseudoscalar meson quantum numbers, for different temperatures 
 are calculated. Analysis is performed
in the Hard Thermal Loop (HTL) Approximation and
leads to a better understanding of the excitations of QGP in
the deconfined phase and is also of relevance for lattice studies.
\end{abstract}

\keywords{Meson screening mass  \*\  Finite temperature QCD  \*\  Quark Gluon Plasma 
 \*\  Meson correlation function 
 \*\  Meson spectral function  \*\  Finite momentum  \*\ HTL approximation.}
\pacs{10.10.Wx,  11.55.Hx,  12.38.Mh,  14.65.Bt,  14.70.Dj,  25.75.Nq}

\maketitle

\section{Introduction}

In this paper we consider the scalar and pseudoscalar mesonic correlation function at high temperature Quantum 
Chromodynamics (QCD) in the framework of the Hard Thermal Loop (HTL) approach. The evaluation of the mesonic correlator 
at finite momenta allows, by Fourier transform, to get information on its large distance behavior. This, in turn, 
is governed by the mesonic screening mass, a quantity which has been evaluated in lattice QCD at large temperature. 
Hence the present, perturbative results can be compared and tested versus the lattice results.

A similar approach was carried out by the authors a few years ago for the pseudoscalar channel only\cite{aga}. The results 
were encouraging but showed some non-negligible discrepancy with respect to the lattice data. The present work contains 
significant improvements in the numerical precision of the calculations as well as a new approach to the analytical 
representation of the spatial correlators, thus providing a more accurate extraction of the mesonic screening masses 
in the QGP. Moreover the investigation is extended to the scalar channel, for which more recent lattice QCD data are available, 
hence allowing for a more complete discussion of the present result.

The features of the meson-like excitations inside the hot Quark-Gluon Plasma (QGP) provide interesting information on the 
persistence of interaction effects up to rather large temperatures, a characteristic which renders QGP a somewhat peculiar 
status of matter and strengthens the interest on the present experiments which aim to investigate and clarify the many issues 
and questions posed by QCD calculations.

In spite of the well-grounded reputation of lattice QCD results, an analytical calculation, like the one presented in this work, 
although contained within the limits of the model approach, may allow a deeper understanding of the physical behavior of quarks 
inside the plasma and the identification of the relevant degrees of freedom. 

The paper is organized as follows: in Section 2 we shortly recall the details of the calculation of the mesonic spatial 
correlation function and spectral function, both within the HTL approximation and for the free case. In Section 3 we 
illustrate the details of the fitting procedure adopted in order to derive precise values of the asymptotic mass. The 
results for the scalar and pseudoscalar channels are shown and compared with a few lattice data. Comments and conclusions are 
reported in Section 4.

\section{Scalar meson spatial correlation function}\label{sec:mesons}

The spatial correlation function is conveniently obtained from the finite temperature correlator of currents carrying 
the proper quantum numbers to create/destroy mesons:
\begin{equation}
{\rm G}_M(-i\tau,\bold{x})\equiv\langle\wtilde{J}_M(-i\tau,\bold{x})
\wtilde{J}_M^{\dagger}(0,\bold{0})\rangle\,,
\label{eq:corr1}
\end{equation}
where $\tau\in[0,\beta=1/T]$ and $\tilde J_M$ denotes the  fluctuation of the current operator $J_M=\bar q \;
 \Gamma_M q$,
the vertex $\Gamma_M$ selecting the appropriate channel (scalar, pseudoscalar, etc.).
The correlator in Eq.(\ref{eq:corr1}) is usually expressed through its Fourier components:
\begin{equation}
 {\rm G}_M(-i\tau,\bold{x})= 
 \frac{1}{\beta}\sum_{n=-\infty}^{+\infty}
 \int\frac{d^3p}{(2\pi)^3}e^{-i\omega_n\tau}
e^{i\bold{p}\cdot\bold{x}}G_M(i\omega_n,\bold{p})\;,
\label{eq:corr2}
\end{equation}
$\omega_n=2n\pi T$ ($n=0,\pm1,\pm2\dots$) being the bosonic Matsubara frequencies.

We are now interested in the $z$-axis correlation function $\mathcal{G}(z)$,
\begin{equation}
 \mathcal{G}_M(z)\equiv\int\limits_0^\beta d\tau \int d\bold{x_\perp}
 {\rm G}_M(-i\tau,\bold{x_\perp},z)\;,
\end{equation}
where the integrations select in Fourier space the vanishing frequency and transverse momentum components.
The asymptotic behavior of static plane-like perturbations with the quantum numbers of a (scalar) meson 
is expected to be exponentially suppressed at large values of $z$ (the distance from the plane in the 
transverse direction)
\beq
\mathcal{G}(z)\underset{z\to +\infty}{\sim}e^{\displaystyle{-m_{scr}z}}\;.
\eeq
The suppression parameter, which is governing the large distance behavior, is the 
so-called \emph{screening mass} $m_{scr}$. It gives information on the nature of the 
 excitations characterizing the QGP phase; the correlator $\mathcal{G}(z)$ is frequently studied 
in lattice simulations, from which the screening mass can be extracted as well.

One can express the spatial correlation function through the finite momentum
spectral function of the quark-antiquark excitations in QGP, $\sigma(\omega,\bold{p})$. Here we shall
consider explicit formulas for the scalar channel only, though results will be presented both for the scalar 
and pseudoscalar channels. The latter was dealt with in Refs \cite{aga,pc} and we refer the reader to 
these works; moreover in the massless case chiral symmetry provides similar formulas for the scalar and pseudoscalar cases, 
as it will be pointed out below.

In the scalar channel the spectral function $\sigma_{S}(\omega,\bold{p})$ is related to $\mathcal{G}_S(z)$ by:
\beqa 
\mathcal{G}_S(z) & = & \int\limits_{-\infty}^{+\infty}\frac{dp_z}{2\pi}
e^{\displaystyle{i p_z z}}\mathcal{G}_{S}(p_z)
\nonumber\\
{} & = & \int\limits_{-\infty}^{+\infty}\frac{dp_z}{2\pi}
e^{\displaystyle{i p_z z}}\int_{-\infty}^{+\infty}d\omega
\frac{\sigma_{S}(\omega,\bold{p_\perp}\!\!=\!0,p_z)}{\omega}\;,\label{eq:gzet}
\eeqa
where $p_z$ is the momentum in the $z$-direction. This is the starting point for investigating 
$z$-axis correlations.

 In spectral representation one can relate the spectral function $\sigma_{S}(\omega,\bold{p})$  
with the meson propagator $\chi_{S}(i\omega_n,\bold{p})$ as follows:
\beq
\chi_{S}(i\omega_n,\bold{p})=-\int\limits_{-\infty}^{+\infty}d\omega
\frac{\sigma_{S}(\omega,\bold{p})}
{i\omega_n-\omega} \quad \Rightarrow \quad \sigma_{S}(\omega,\bold{p})=
\frac{1}{\pi}\trm{Im}\,\chi_{S}(\omega+i\eta,\bold{p}).\label{eq:specrit}
\eeq

The scalar meson 2-point function \cite{berry}, in turn, is
\beq\label{chi}
\chi_S(i\omega_l,\bold{p})=- N_f N_c \frac{1}{\beta}\!\sum_{n=-\infty}^{+\infty} \int\!\frac{d^3k}
{(2\pi)^3}\trm{Tr}[\id \, S(i\omega_n,\bold{k})\id \, S(i\omega_n\!-i\omega_l,
\bold{k}-\bold{p})]\;,
\eeq
where $S(i\omega_n,\bold{k})$ is the quark propagator and $N_f N_c$ are the numbers
of flavors and colors.

In HTL approximation the quark propagator is dressed by the interaction with the other particles of the 
thermal bath (antiquarks and gluons), carrying typical hard momenta (proportional to the temperature of the 
thermal bath); in the spectral representation one can write:
\beq\label{fermspec}
S^{HTL}(i\omega_{n},\bold{p})=-\int\limits_{-\infty}^{+\infty} d\omega\frac{\displaystyle{\rho_{\trm{HTL}}
(\omega,\bold{p})}}{i\omega_n-\omega}\;,
\eeq
where the HTL quark spectral function is 
\beq\label{spectral}
\rho_{\trm{HTL}}(\omega,\bold{p})=\frac{\gamma^0-i\bold{\gamma\cdot\hat{p}}}{2}\rho_{+}(\omega,p)\, +\, 
\frac{\gamma^0+i\bold{\gamma\cdot\hat{p}}}{2}\rho_{-}(\omega,p) \; ,
\eeq
with
\beq
\rho_{\pm}(\omega,k)=\frac{\omega^2-k^2}{2m_q^2}[\delta(\omega-\omega_{\pm})+\delta(\omega+\omega_{\mp})]
+\beta_{\pm}(\omega,k)\theta(k^2-\omega^2)\;,\label{2part}
\eeq
and
\beq
\beta_{\pm}(\omega,k)=-\frac{m_q^2}{2}\frac{\pm\omega-k}{\left[k(-\omega\pm k)+m_q^2\left(\pm1-\frac{
\pm\omega-k}{2k}\ln\frac{k+\omega}{k-\omega}\right)\right]^2+\left[\frac{\pi}{2}m_q^2\frac{\pm\omega-k}
{k}\right]^2}\;.
\eeq
Here the \emph{thermal gap mass} of the quark is $m_q=gT/\sqrt{6}$. 

Inserting (\ref{fermspec}) into (\ref{chi}) and setting $\bold{q}=\bold{k}-\bold{p}$  one gets:
\begin{multline}
\chi_S^{\trm{HTL}} (i\omega_l,\bold{p})\!=-N_f N_c\frac{1}{\beta}\!\sum_{n=-\infty}^{+\infty}\!\!\int\!\frac{d^3k}
{(2\pi)^3}\int\limits_{-\infty}^{+\infty}d\omega_1\int\limits_{-\infty}^{+\infty}d\omega_2\frac{1}{i\omega_n-\omega_1}
\frac{1}{i\omega_n-i\omega_l-\omega_2} \\
\times\trm{Tr}[\id \rho_{\trm{HTL}}(\omega_1,\bold{k})\id \rho_{\trm{HTL}}(\omega_2,\bold{q})]
\;.\label{eq:mesonHTL}
\end{multline}
Then we sum over the Matsubara frequencies in Eq.~(\ref{eq:mesonHTL}) with a standard contour integration, 
we perform the usual analytical continuation $i\omega_{l}\rightarrow\omega +i\eta^+$ (corresponding to 
retarded boundary conditions) and extract the imaginary part of the result, thus obtaining:
\begin{multline}
\sigma_S^{\trm{HTL}}(\omega,\bold{p})= N_f N_c\int\!\frac{d^3k}{(2\pi)^3}(e^{\beta\omega}-1)
\int\limits_{-\infty}^{+\infty}d\omega_1\int\limits_{-\infty}^{+\infty}d\omega_2\tilde{n}(\omega_1)\tilde{n}(\omega_2) \\
\times\delta(\omega-\omega_1-\omega_2)\cdot\trm{Tr}[\id \rho_{\trm{HTL}}(\omega_1,\bold{k})
\id \rho_{\trm{HTL}}(-\omega_2,\bold{q})]\;.\label{eq:sigmap}
\end{multline}
Now, by inserting Eq. (\ref{spectral}) into Eq. (\ref{eq:sigmap}) and since
\bseq
\begin{align}
\trm{Tr}\left[\id \frac{\gamma^0\mp i\bold{\gamma\cdot\hat{k}}}{2}\id \frac{\gamma^0\mp
i\bold{\gamma\cdot\hat{q}}}{2}\right]&=(1-\bold{\hat{k}\cdot\hat{q}}),\\
\trm{Tr}\left[\id \frac{\gamma^0\mp i \bold{\gamma\cdot\hat{k}}}{2}\id \frac{\gamma^0\pm i
\bold{\gamma\cdot\hat{q}}}{2}\right]&=(1+\bold{\hat{k}\cdot\hat{q}})\;,
\end{align}
\eseq
one gets 
\begin{multline}
\sigma_S^{\trm{HTL}} (\omega,\bold{p})\!=N_f N_c\int\!\frac{d^3k}{(2\pi)^3}(e^{\beta\omega}-1)
\int\limits_{-\infty}^{+\infty}d\omega_1\int\limits_{-\infty}^{+\infty}d\omega_2\tilde{n}(\omega_1)\tilde{n}
(\omega_2)\delta(\omega-\!\omega_1-\!\omega_2)\\
\times\left\{(1+\bold{\hat{k}\cdot\hat{q}})[\rho_+(\omega_1,k)\rho_+(\omega_2,q)+\rho_-(\omega_1,k)
\rho_-(\omega_2,q)] \right.\\
+\left.(1-\bold{\hat{k}\cdot\hat{q}})[\rho_+(\omega_1,k)\rho_-(\omega_2,q)+\rho_-(\omega_1,k)
\rho_+(\omega_2,q)]\right\}\;,\label{eq:sigmapcompl}
\end{multline}
which is exactly equal to the $\sigma_{\trm{PS}}(\omega,\bold{p})$ of Ref.~\cite{pc}, since for massless
quarks chiral symmetry holds (the thermal mass $m_q$ acquired in the bath does not affect the symmetries 
of the original Lagrangian).

In the case of a non-interacting system of quarks with mass $m$, at finite momentum the free quark 
spectral function reads:
\beq
\rho^{free}(K)=(K\hspace{-.25cm}{\slash}+m)\frac{1}{2\epsilon_k}
[\delta(k_0-\epsilon_k)-\delta(k_0+\epsilon_k)]\,,
\label{freero}
\eeq
where $\epsilon_k = \sqrt{k^2+m^2}$ and $K=(k_0,\bold{k})$. Then, by inserting [instead of $\rho^{HTL}$] 
(\ref{freero}) into Eq.~(\ref{eq:sigmap}) one gets the analytic expression~\cite{aar}:
\begin{multline}\label{eq:freefinite}
\sigma^{free}_S(\omega,\bold{p})\!=\! \frac{N_c N_f}{8
\pi^2} (\omega^2 - p^2-4 m^2)\\
\times\left\{\theta(\omega^2-p^2-4 m^2)\left[\sqrt{1-\frac{4 m^2}{\omega^2-p^2}}+
\frac{2}{p \beta}A\right]+\theta(p^2-\omega^2)\frac{2}{p \beta} B \right\} ,
\end{multline}
where 
\beqa
A = \log\left(1+e^{-\frac{\beta}{2}\Big(\omega+p\sqrt{1-\frac{4 m^2}{\omega^2-p^2}}\Big)}\right)
\!-\!\log\left(1+e^{-\frac{\beta}{2}\Big(\omega-p\sqrt{1-\frac{4 m^2}{\omega^2-p^2}}\Big)}\right), \nonumber
\\
\nonumber
B = \log\left(1+e^{-\frac{\beta}{2}\Big(\omega+p\sqrt{1-\frac{4 m^2}{\omega^2-p^2}}\Big)}\right)\!
-\!\log\left(1+e^{+\frac{\beta}{2}\Big(\omega-p\sqrt{1-\frac{4 m^2}{\omega^2-p^2}}\Big)}\right) .
\eeqa
For high energy ($\omega \to \infty$) the spectral function diverges quadratically  %($\sigma(\omega,p_z) \to \omega^2$), 
($\sigma \to \omega^2$), hence in order to get a finite result for $\mathcal{G}(z)$ Eq.~(\ref{eq:gzet}) one has to
regularize integrals. For the non interacting QGP the problem was solved for all mesonic channels in Ref.~\cite{piotr} 
by adopting the Pauli-Villars regularization scheme. In the scalar channel one gets:
\beqa 
\mathcal{G}^{free}_S(z) & = & \frac{N_f N_c T}{4 \pi z^2}  
 \label{eq:free}\\ 
{} & \times &   \sum_{l=-\infty}^{+\infty} 
e^{\displaystyle{-2 z} \sqrt{(2 l+1)^2\pi^2T^2+m^2}}
\left( 2 z  \sqrt{(2 l+1)^2\pi^2T^2+m^2} +1 \right),\nonumber 
\eeqa
The same procedure cannot be adopted for the interacting case, since the meson spectral function is obtained only via 
numerical evaluation. Nevertheless a ``numerical'' regularization of the integrals can be performed by following the 
same method proposed in Ref.~\cite{aga}. We notice that the high frequency divergence is of the same order both for the free and 
for the interacting system; hence we first define the difference between the two spectral functions: 
\beq 
\sigma_{S}^{diff}(\omega, p_z) = \sigma_{S}^{free}(\omega, p_z) -
\sigma_{S}(\omega, p_z).  \label{eq:diff}
\eeq
The numerical evaluation of this quantity was accurately tested to converge to zero for high energies and, 
for a given momentum $p_z$, it can be numerically integrated as well. In Eq.~(\ref{eq:diff}) the asymptotic 
mass $m = \sqrt{2} m_q$ has been used in the free spectral function (the so-called auxiliary 
spectral function of Ref.~\cite{aga}).

\begin{figure}[!htp]
\begin{center}
\includegraphics[clip,width=0.8\textwidth]{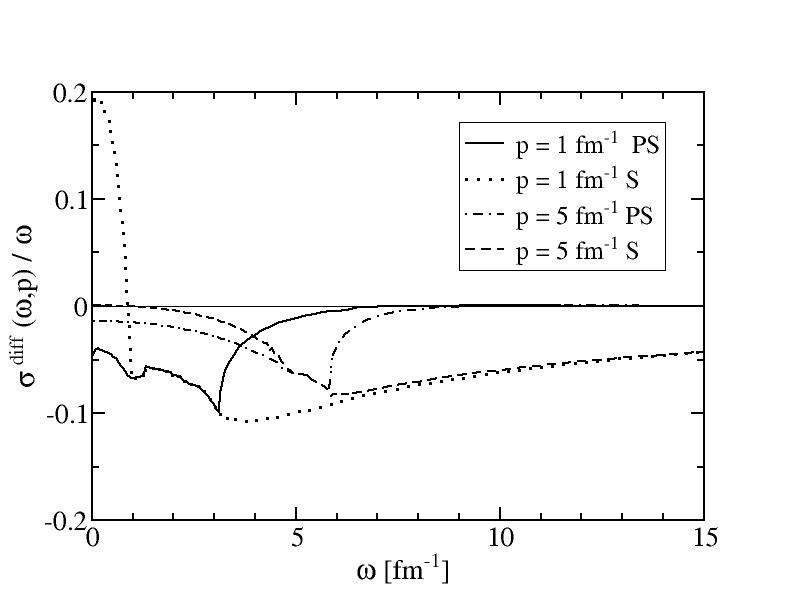}\\
\includegraphics[clip,width=0.8\textwidth]{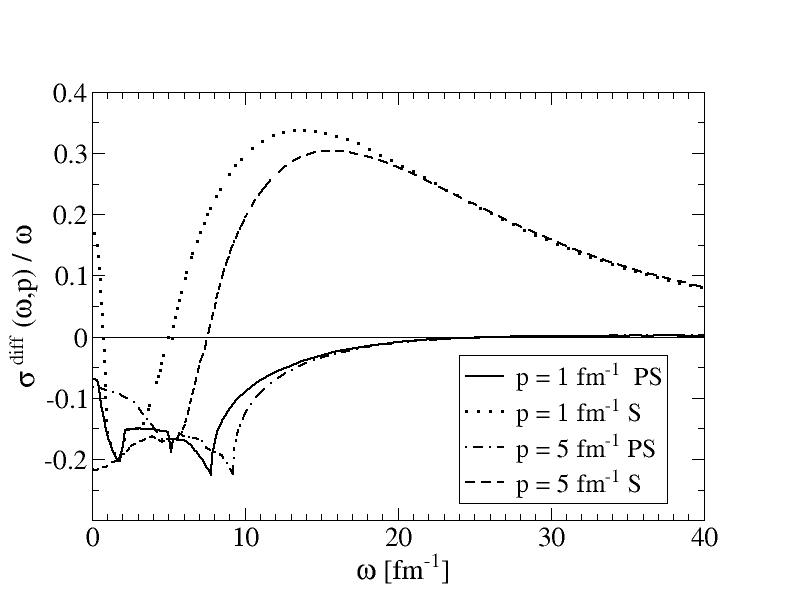}
\caption{The meson scalar and pseudoscalar differences of spectral functions,  
Eq.~(\ref{eq:diff}) divided by  $\omega$, for different momenta $p=p_z$ and different temperatures: 
$T/T_c=1$ in the upper panel, $T/T_c=4$ in the lower panel.}\label{p06} 
\end{center}
\end{figure}

Here we shall not limit ourselves to consider the scalar channel, for which no result exists yet: 
we also present results for the pseudoscalar channel since we have significantly increased the 
precision of the numerical calculations with respect to previous works~\cite{aga}. 

 In Fig.~\ref{p06} we show the $\omega$ dependence of Eq.~(\ref{eq:diff}) for a few values of the momenta $p_z$.
The numerical calculations were carried out up to  $\omega=2500 fm^{-1}$, in order to check the perfect convergence to zero, 
in the infinite $\omega$ limit, of the difference between the full and the free spectral functions.

Then we fit the numerically obtained $\mathcal{G}^{diff}_S(p_z)=\mathcal{G}^{free}_S(p_z)-\mathcal{G}_S(p_z)$ 
to a sum of Yukawa like functions:
\beq 
\mathcal{G}^{diff}(p_z) =  \sum_{i=1}^{n} \frac{2 m_i c_i}{m_i^2+p_z^2}
\;,\label{eq:fit}
\eeq
which can be easily converted by the Inverse Fourier Transform (IFT) to ${G}^{diff}(z)$ in coordinate space
\beq 
\mathcal{G}^{diff}(z) = \int\limits_{-\infty}^{+\infty}\frac{dp_z}{2\pi}\; \;
e^{\displaystyle{i p_z z}}\; \;\mathcal{G}^{diff}(p_z) =  \sum_{i=1}^{n} c_i  e^{-m_i  z }\;.\label{eq:gzdif}
\eeq

The parameters $m_i$ and $c_i$ for different temperatures ranging from 1 to $10~T_c$ are collected in 
Table~\ref{tab1} for the pseudo-scalar channel and in Table~\ref{tab1a} for the scalar one. We remark that
two terms are sufficient in the pseudoscalar channel, while the scalar one requires up to four terms, 
depending upon the temperature. The number of digits reported in the tables is required to ensure an accurate 
fit of the function.

%
% table 1
\begin{table}[h]
\begin{center}
\begin{tabular}{r|rrrr}
$T/T_c$ & $c_1$ & $m_1$ & $c_2$ & $m_2$\\
\hline \hline
1 & -0.7159 & 6.7027  & -23.5261 &  148.9711  \\
2 & -3.5432 & 13.5179  & -106.3161 & 285.0631  \\
4 & -17.7817 & 24.6371  & -290.4082 & 251.5351  \\
10 & -202.8622 & 60.8533  & -1931.3612 & 355.4231  \\
\end{tabular}
\vspace{0.3cm}
\caption{\label{tab1}Parameters of $\mathcal{G}^{diff}(p_z)$ defined in Eq.~(\ref{eq:fit}) for the pseudoscalar channel; 
the $m_i$ are in [fm$^{-1}$].}
\end{center}  
\end{table}
% table 2
\begin{table}
\begin{center}
\begin{tabular}{r|rrrrrrrr}
$T/T_c$ & $c_1$ & $m_1$ & $c_2$ & $m_2$ & $c_3$ & $m_3$ & $c_4$ & $m_4$\\
\hline \hline
1 & -92.31 & 84.98 & -834.71 & 512.86 & -17.92 & 17.52 & -1.92 & 5.34 \\
2 & -44.95 & 15.86 & 30.08 & 8.42 & -323.78 & 317.84 & & \\
4 & 104.97 & 10.27 & 1088.71 & 487.25 & & & &\\
10 & 1176.17 & 231.28 & 1084.73 & 30.03 & 1081.82 & 30.05 & 88.52 &10.54\\
\end{tabular}
\vspace{0.3cm}
\caption{\label{tab1a}Parameters of $\mathcal{G}^{diff}(p_z)$ defined in Eq.~(\ref{eq:fit}) for the scalar channel; 
the $m_i$ are in [fm$^{-1}$].}
\end{center}  
\end{table}

\begin{figure}[!htp]
\begin{center}
\includegraphics[clip,width=0.49\textwidth]{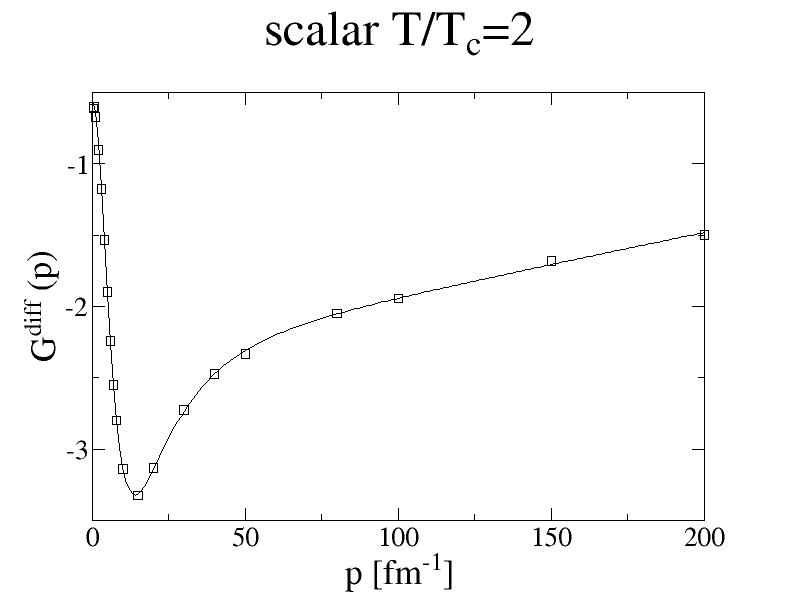}
\includegraphics[clip,width=0.49\textwidth]{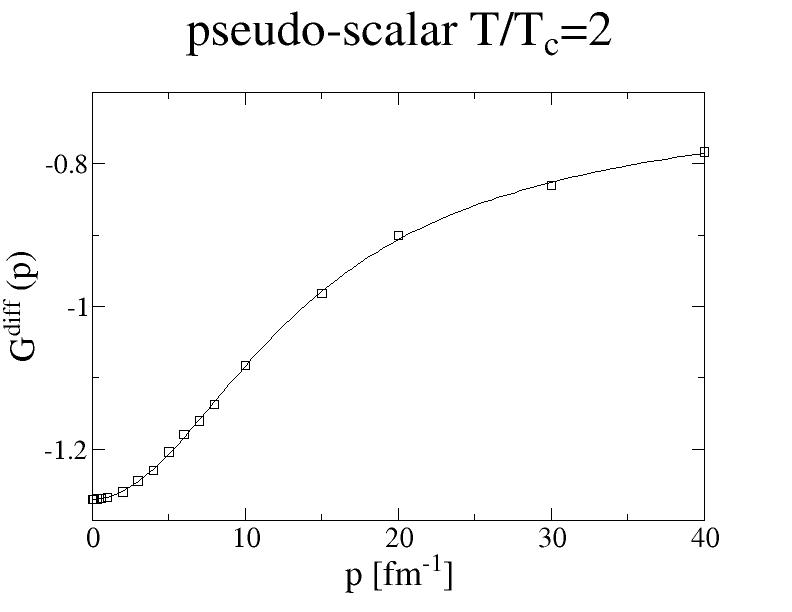}\\
\includegraphics[clip,width=0.49\textwidth]{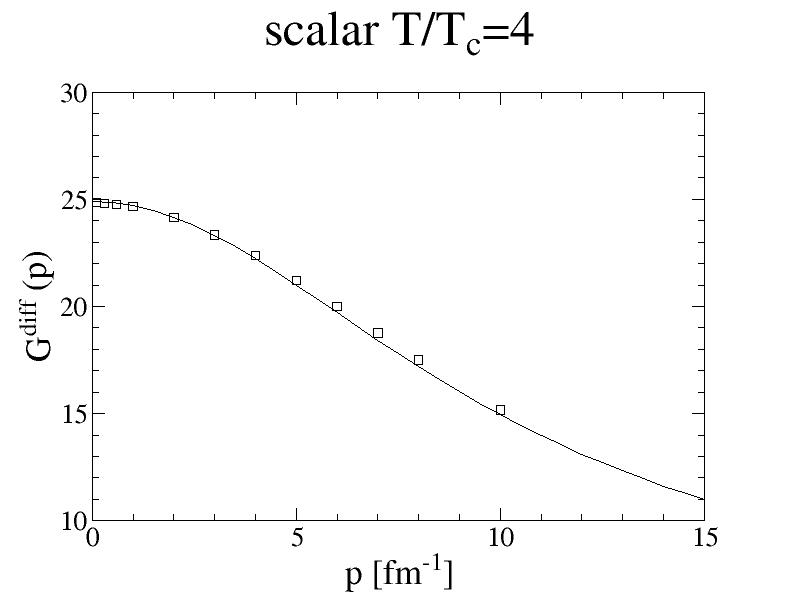}
\includegraphics[clip,width=0.49\textwidth]{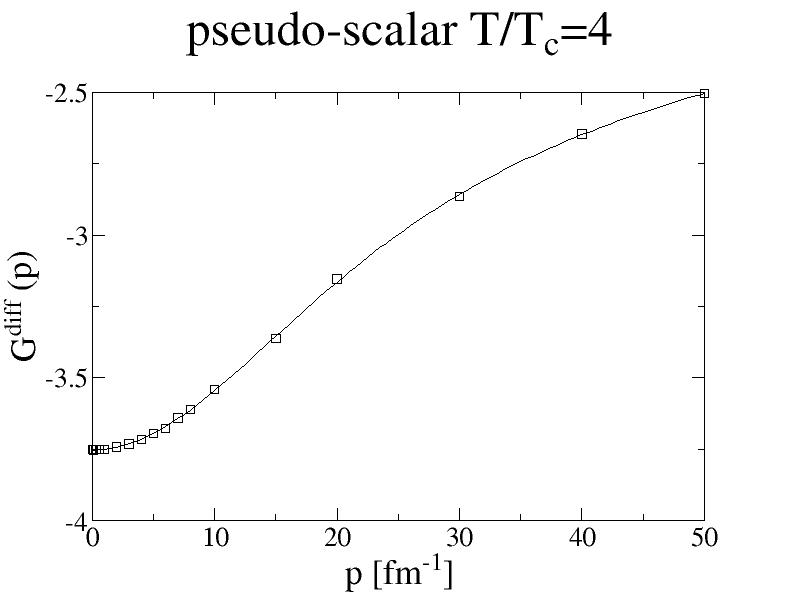}\\
\caption{Fit for the meson scalar and pseudoscalar difference, Eq.(\ref{eq:fit}) of the momentum $p=p_z$
correlation function as a function of $\omega$, for two values of $T/T_c$.}\label{gp} 
\end{center}
\end{figure}

The accuracy of the fit is shown in Fig.~\ref{gp} for the two channels, at two different
temperatures above $T_c$. One can notice a similar behavior (apart from sign and size) at the higher 
temperature, while $T/T_c=2$ shows that the two channels have the same sign, but the minimum in 
the scalar channel is absent in the pseudoscalar one.

After Fourier transforming, according to Eq.~(\ref{eq:gzdif}), the final formula for the full spatial 
correlation function in coordinate space is:
\beq 
\mathcal{G}^{full}(z) = \mathcal{G}^{free}(z) - \mathcal{G}^{diff}(z) 
\;.\label{eq:ful}
\eeq
For large $z$  the final results (\ref{eq:ful}) is once again fitted to an analytic function as follows
\beq 
\mathcal{G}^{full}(z) =  \sum_{i=1}^{n} b_i  \frac{e^{-m_i  z }}{z}\;.\label{eq:gz}
\eeq

At variance with the previous approach, where we have assumed that the interaction part ${G}^{diff}(p_z)$
of the correlation function is dominating and we have extracted the screening mass $m_{scr}= m_1$ directly 
from its asymptotic contribution, namely from the lightest masses of Table~\ref{tab1} (only the pseudoscalar 
channel was considered in Ref. \cite{aga}), here we improve the fitting procedure by extracting the screening 
masses directly from the fit of the the total spatial correlation function Eq.~(\ref{eq:gz}). This makes a
small but significant difference (in the previously considered channel): indeed now the masses  approach the 
high temperature limit from below the free, massless non interacting system limit ($2\pi T$). We should also notice that, while 
in the pseudoscalar channel the choice of the lightest mass is rather obvious at all considered temperatures, 
in the scalar channel this procedure would be questionable (see Table~\ref{tab1a}).

\begin{figure}[!htp]
\begin{center}
\includegraphics[clip,width=0.49\textwidth]{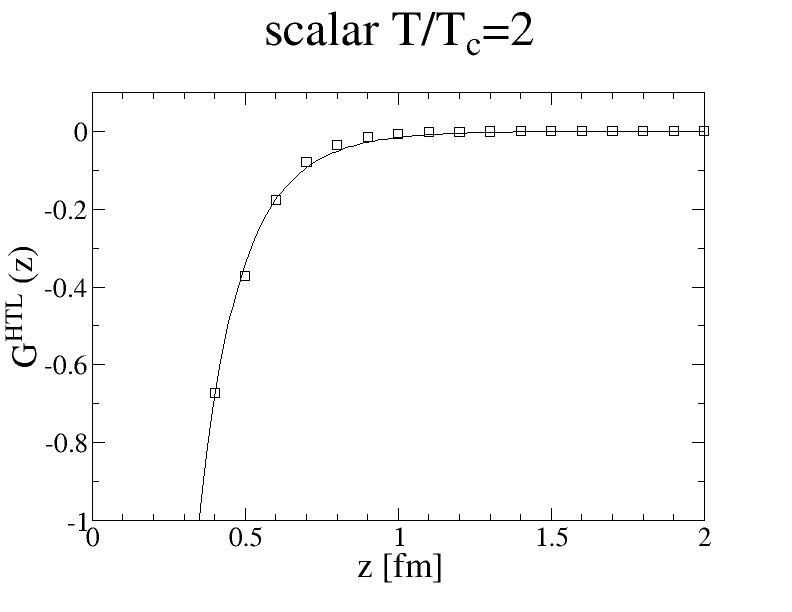}
\includegraphics[clip,width=0.49\textwidth]{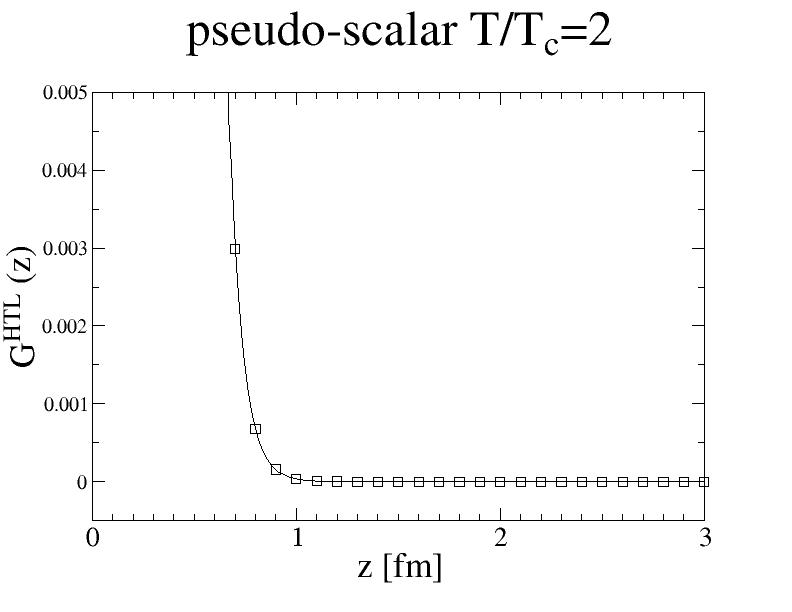}\\
\includegraphics[clip,width=0.49\textwidth]{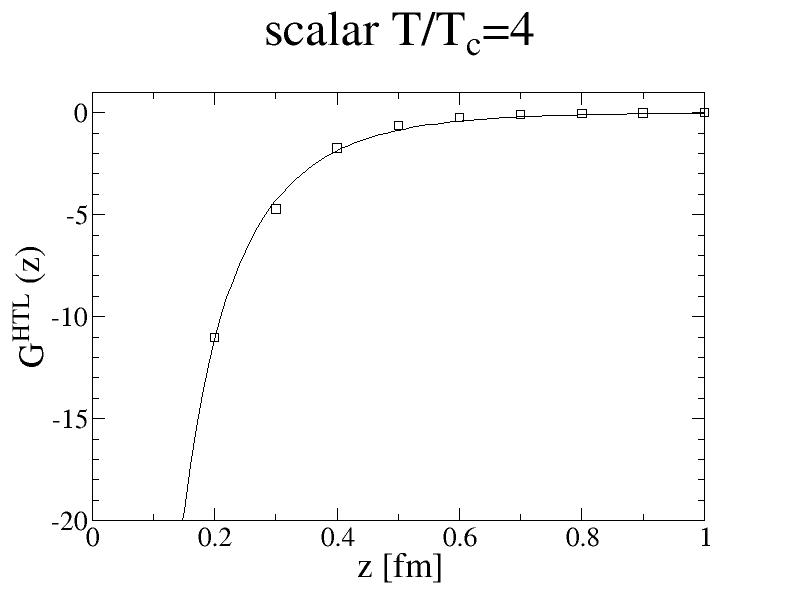}
\includegraphics[clip,width=0.49\textwidth]{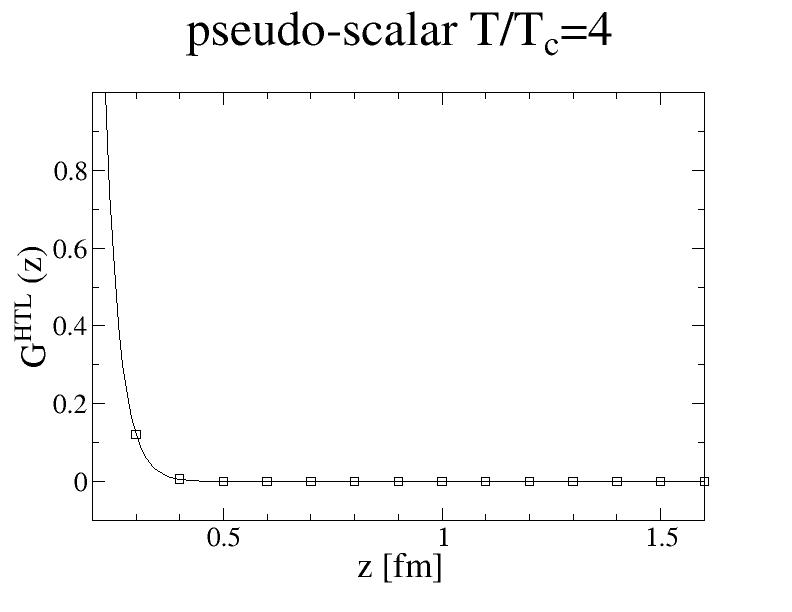}\\
\caption{Fit for the meson scalar and pseudoscalar (\ref{eq:gz})  spatial
correlation function for different $T/T_c$ values.}\label{gz} 
\end{center}
\end{figure}

% table 3
\begin{table}
\begin{center}
\begin{tabular}{r|rrrr}
$T/T_c$ & $b_1$ & $m_1$ & $b_2$ & $m_2$\\
\hline \hline
1 & 6.843 & 6.953 & 2.381 $10^{-2}$& 4.875  \\
2 & 26.643 & 13.505 & 8.133 $10^{-3}$ & 10.466 \\
4 & 101.705 & 26.389 & -5.991 $10^{-1}$ & 23.027  \\
10 & 626.345 & 65.553 &  &   \\
\end{tabular}
\caption{\label{tab2}Parameters of $\mathcal{G}^{full}(z)$ defined in Eq.~(\ref{eq:fit})
for the pseudoscalar channel; the $m_i$ are in [fm$^{-1}$].}
\end{center}  
\end{table}
%
% table 4
\begin{table}
\begin{center}
\begin{tabular}{r|rrrr}
$T/T_c$& $b_1$ & $m_1$ & $b_2$ & $m_2$ \\
\hline \hline
1 & 5.8132 & 6.9986 & 0.8941 & 4.5283 \\
2 & -18.2976 & 9.5012 & & \\
4 & -66.1941 & 21.7911 & &\\
10 & -166.7731 & 75.5921 & 328.4512 & 64.2514\\
\end{tabular}
\caption{\label{tab2a}Parameters of $\mathcal{G}^{full}(z)$ defined in Eq.~(\ref{eq:fit}) 
for the scalar channel; the $m_i$ are in [fm$^{-1}$].}
\end{center}  
\end{table}

In Table~\ref{tabm} we summarize the results for the screening masses at all temperatures 
and for all channels considered here. The numbers in the table are calculated assuming
$N_f=2$ and $T_c=202$~MeV for the transition temperature, according to Ref.~\cite{kacz} 
(this value is also close to the one resulting from the lattice calculations we shall compare later on). 
The ratio $T_c/\Lambda_{\overline{MS}} = 0.7721$ for the $N_f=2$ case was taken from 
Refs.~\cite{karsch,goc,zan}. Throughout the calculation we have employed a running gauge 
coupling as given by the two-loop perturbative beta-function, leading to the expression:
\beq
g^{-2}(T) = 2 b_0 \log{\frac{\mu}{\Lambda_{\overline{MS}}}} +
\frac{b_1}{b_0} \log\Big\{2 \log{\frac{\mu}{\Lambda_{\overline{MS}}}}
\Big\}  , \label{eq:running}
\eeq
where
$
b_0=\frac{1}{16 \pi^2} \Big( 11 - 2 \frac{N_f}{3} \Big)$, $
b_1=\frac{1}{(16 \pi^2)^2} \Big( 102 - 38 \frac{N_f}{3} \Big)
$.
The choice of the renormalization scale $\mu$ should reflect the typical 
momentum exchanged by the particles which, in an ultra-relativistic plasma, 
is of order $T$. Here, for the sake of simplicity, we adopt the choice $\mu = 1.142 \pi T$, 
which was suggested in Ref.~\cite{kacz}. 
For reference, we also recall that the thermal gap mass of the quark is $m_q=g(T) T/\sqrt{6}$.\\

% table 5
\begin{table}
\begin{center}
\begin{tabular}{r|rrrr}
$T/T_c$  & $ m /2\pi T$  & $m_{scr}^{free}$/$2\pi T$ & $m^{scr}_{PS}$/$2\pi T$ & $m^{scr}_{S}$/$2\pi T$\\
\hline \hline
1 & 0.229  & 1.101 & 0.758 & 0.704  \\
2  & 0.176  & 1.060 & 0.814 & 0.738 \\
4  & 0.150  & 1.044 & 0.895 & 0.847 \\
10  & 0.129  & 1.033 & 1.019 & 0.998 \\
\end{tabular}
\caption{\label{tabm}
 The columns display, respectively: $m = \sqrt{2} m_q$, the HTL asymptotic quark mass, 
 $m_{scr}^{free}= 2 \sqrt{\pi^2T^2+m^2}$, the free screening mass, 
 $m^{scr}_{PS}$ and   $m^{scr}_{S}$, which are the screening masses of the interacting QGP, in the pseudo-scalar and scalar channels, respectively}
\end{center}  
\end{table}

\begin{figure}[!htp]
\begin{center}
\includegraphics[clip,width=0.49\textwidth]{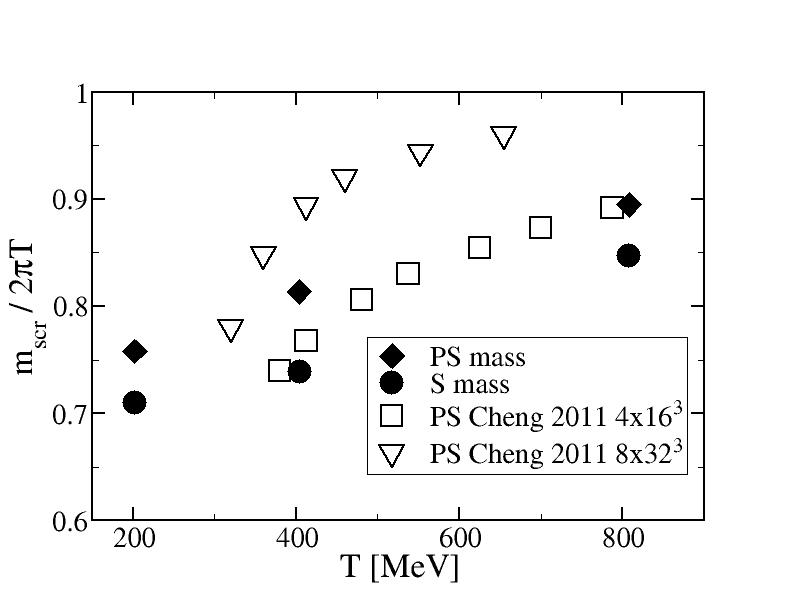}
\includegraphics[clip,width=0.49\textwidth]{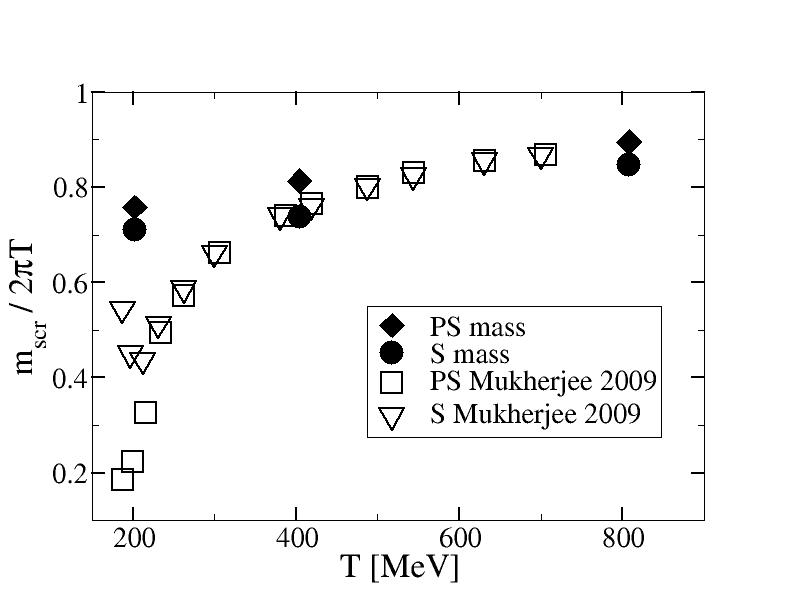}\\
\caption{Temperature dependence of the scalar and pseudoscalar screening masses
 compared with the lattice
results extracted from Refs \cite{chang,muk9}}\label{masy} 
\end{center}
\end{figure}

Next we compare our results for the screening masses with recent lattice results. In particular we refer
to the lattice data of Cheng {\it et al.}~\cite{chang}, which are the most recent and employ two different 
lattice sizes, but are obtained in pseudoscalar channel only; in addition we show a comparison with the 
results of Mukherjee~\cite{muk9}, who also provides data in the scalar channel.
Fig.~\ref{masy} shows the temperature dependence of the screening masses obtained in the present work, 
both in the scalar (black circles) and pseudoscalar (black diamonds) channels, divided by the free massless
asymptotic limit ($2\pi T$). Three values of temperature are shown (1, 2 and 4~$T_c$), the last being the upper 
limit for which lattice data are also available. We report our result at $T=T_c$ only for sake of comparison, 
but we remind the reader that the HTL approach can be considered as a reliable one only at higher temperatures 
(e.g. $T=2T_c$ or more).

In the left panel we compare with the results of Cheng {\it et al.}~\cite{chang} in the pseudoscalar channel. 
The agreement is not very satisfactory and appears to be better for the smaller lattice (open squares). 
In the right panel we compare  with the results of Mukherjee~\cite{muk9}, which allow for a full comparison in 
both channels: with the exception of the lowest temperature points, already commented above, the agreement 
between lattice data and our calculation is rather good. The scalar and pseudoscalar lattice results are closer
to each other than in our approach, but the trend and the size are quite comparable.

\begin{figure}[!htp]
\begin{center}
\includegraphics[clip,width=0.8\textwidth]{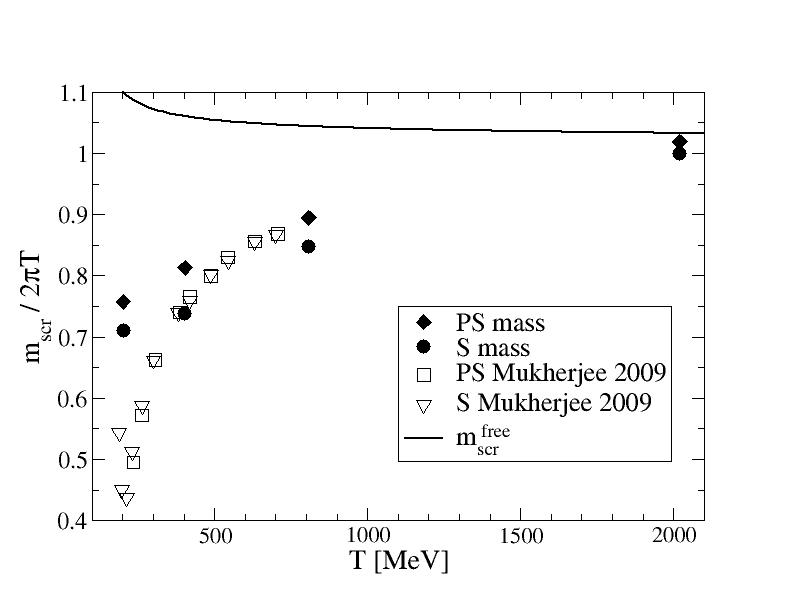}
\caption{Temperature dependence of the scalar and pseudoscalar screening masses
 compared with the lattice results extracted from Ref.~\cite{muk9} and with the screening mass
 in the non-interacting system, $m_{scr}^{free}= 2 \sqrt{\pi^2T^2+m^2}$.}\label{masy10} 
\end{center}
\end{figure}

The same considerations apply to the results illustrated in Fig. ~\ref{masy10}, where we extend the scale of 
the temperature up to $10~T_c$, the highest value considered in this work: here we also compare with the 
screening mass of the ``non-interacting quarks'', namely with the screening mass value of the auxiliary 
spectral function employed to evaluate the full HTL results: $m_{scr}^{free}= 2 \sqrt{\pi^2T^2+m^2}$. The latter
is obviously always larger than $2\pi T$ but we can assume it as the asymptotic value, which in the figure appears
to be almost reached at  $T=10~T_c$.

\section{Conclusions}\label{sec:concl}
The main goal of this paper was to calculate the asymptotic mass, at different temperatures, of the scalar 
and pseudo-scalar mesons in an interacting quark gluon plasma described within the HTL approximation.
For this purpose we considered the scalar and pseudo-scalar mesonic correlation function at high temperature QCD.
The evaluation of the mesonic correlators allowed us to obtain information on its large distance behavior, which, in turn, 
is governed by the mesonic screening mass.

The main issue of the present approach, which extends to the scalar channel the methodology employed in a 
previous work~\cite{aga}, was to improve the precision in the numerical determination of the correlators at
large distances, hence obtaining a more precise determination of the screening mass, the parameter which governs
the large distance behavior. We have thus modified the procedure for fitting the numerically determined HTL 
correlator: the full (interacting) correlator is determined via the asymptotic (analytic) behavior of the free 
system, which allows for an appropriate regularization of the otherwise divergent integrations. Then a sequence 
of fitting procedures, first in momentum space and then in coordinate space, allowed us to determine the 
screening mass more accurately than in the past. 

Although the differences do not appear to be dramatic, an important feature of the screening masses presently obtained
is that they approach, with increasing temperature, the non interacting limit from below rather than from above, as 
in the previous work. The comparison with the available lattice data confirms this characteristic. Moreover the
values for the screening masses we obtained here are in fair agreement with at least one set of lattice data~\cite{muk9}.
Only close to the critical temperature our approach shows disagreement with respect to lattice results, which does not 
come as a surprise, since the HTL approach is considered to be reliable only well above $T_c$.

We also notice that other perturbative computations of the static correlation lengths~\cite{lai,vep}, led to
a small but positive correction to the free screening mass:
\beq
m_{\rm scr}\simeq 2\pi T+\left(\frac{1}{3\pi}+\Delta\right)g^2T\;,
\label{eq:lai}
\eeq
$\Delta$ being an additional correction term, to the same order. The numerical values turn out to be above free limit $2 \pi T$ , whereas
all available results from lattice collaborations lie below it~\cite{kacz,karsch,goc,chang,muk9,taro,taro2,petr,wis,gavai,ban,ban1,kacz1,karsch1,karsch2}.
This fact strengthens the validity of the present approach and particularly of the double fitting procedure which allowed us to obtain
smaller screening masses within, basically, the same HTL approach used in the past.

\section*{Acknowledgments}
One of the authors (P.C.) thanks the Department of Theoretical Physics
 of the Torino University for the warm hospitality in the final phase of this work.


\begin{thebibliography}{99}
\bibitem{aga} W.M. Alberico, A. Beraudo, A.Czerska, P. Czerski, A. Molinari, Nucl. Phys. { A792}, 152 (2007)
\bibitem{pc} W.M. Alberico, A. Beraudo, P. Czerski, A. Molinari, Nucl. Phys. {A775}, 188 (2006)
\bibitem{berry} W.M. Alberico, A. Beraudo, A. Molinari, Nucl. Phys. {A750}, 359 (2005)
\bibitem{aar} G. Aarts, J.M. Martinez Resco, Nucl. Phys. {B726}, 93 (2005)
\bibitem{piotr} P. Czerski, Cent. Eur. J. Phys. {10}, 342 (2012) 	
\bibitem{kacz} O. Kaczmarek, F. Zantow, Phys. Rev. {D71}, 114510 (2005)
\bibitem{karsch} F. Karsch, E. Laermann, A. Peikert, Phys. Lett. {B478}, 447 (2000) 
\bibitem{goc} M. Gockeler {et al.}, Phys. Rev. {D73}, 014513 (2006)
\bibitem{zan} O. Kaczmarek, F. Zantow,  hep-lat/0512031
\bibitem{chang} M. Cheng { et al.}, Eur. Phys. J.{C71}, 1564 (2011) 
\bibitem{muk9} S. Mukherjee, Nucl. Phys. {A820}, 283C (2009) 
\bibitem{lai} M. Laine,  M. Vepsalainen, JHEP 0402, 004 (2004) 
\bibitem{vep} M. Vepsalainen, JHEP 0703, 022 (2007)
\bibitem{taro} P. de Forcrand et al. (QCD-TARO Collaboration), hep-lat/9901017
\bibitem{taro2} I. Pushkina et al. (QCD-TARO Collaboration), Phys. Lett. {B609}, 265 (2005)
\bibitem{petr} P. Petreczky, J. Phys. {G30}, S431 (2004)
\bibitem{wis} S. Wissel et al., PoS LAT2005, 164 (2006)
\bibitem{gavai} R. V. Gavai, S. Gupta, R. Lacaze, Phys. Rev. {D78}, 014502 (2008)
\bibitem{ban} D. Banerjee, R. V. Gavai, Sourendu Gupta, Phys. Rev. {D83}, 074510 (2011)
\bibitem{ban1} D. Banerjee, R. V. Gavai, Sourendu Gupta, PoSLAT2010, 168 (2010) 
\bibitem{kacz1} O. Kaczmarek,  PoSCPOD07, 043 (2007) 
\bibitem{karsch1} F. Karsch, PoSCPOD07,  026 (2007) 
\bibitem{karsch2} F. Karsch, PoSLAT2007, 015 (2007)
\end{thebibliography}
\end{document}